\documentstyle[prl,aps,amsfonts,amssymb,twocolumn,epsfig]{revtex} 
\begin{document}

\preprint{
\hfill  TH99.2
} 

\title{
Manifestation of intermediate spin for Fe$_{8}$.
} 

\author{S. E. Barnes}

\address{Department of Physics, University of Miami, Coral Gables, 
Florida 33124 }

% 
% \draft

% \twocolumn[%caution - le "[" ici et en bas est essential
% 
% \hsize\textwidth\columnwidth\hsize\csname @twocolumnfalse\endcsname

\date{\today} \maketitle
\begin{abstract}
Intermediate spin, which occurs in the theory of anyons, can also be 
exhibited by mesoscopic magnetic particles.  The necessary broken 
time reversal symmetry is due to a suitably directed magnetic field.  
As a function of this field a system passes periodically through 
points which correspond to whole or half-integer spin.  Intermediate 
spin is defined by fields which lie between these points.  Since the 
tunnel splitting in the ground doublet vanishes for half-integer spin, 
this splitting becomes periodic.  The doubly periodic oscillations 
observed in the magnetic molecular cluster Fe$_{8}$ represent the 
first unequivocal observation of this phenomena.  Here the whole or 
half-integer nature of the system, i.e., the parity, is periodic for a 
field which is along either the easy or hard axis or a suitable 
combination of the two.  A detailed theory is presented.

\end{abstract}

\pacs{ 73.40.Gk, 75.60.Jp, 75.10.Jm, 03.65.Sq, 75.30.Gw}

% \vskip1.0pc]%caution - le "[" est essential

The search for quantum effects in mesoscopic systems has become quite 
fashionable.  In particular, relatively recently advances in several 
technologies have encouraged the investigation of the cross-over 
between the classical and quantum regimes in magnetic molecular 
clusters.  A particularly interesting quantum effect which can occur 
in such large spin clusters has been emphasized by Loss et al. \cite{refI} 
and von Delft and Henley \cite{refII}.  For half-integer spin the amplitude for 
tunneling with a complete reversal of the magnetization of a small 
ferromagnet is zero due to interference effects arising from the so 
called {\it topological term\/} in the effective action.  This parity 
effect implies that, while a whole integer spin system can have a tunnel 
split ground state doublet, the half-integer analogue will not.  Garg 
\cite{refIII} has shown, for a small easy axis ferromagnet with a field 
perpendicular to the direction of the magnetization, the same 
tunneling amplitude oscillates as a function of field.

\def\refI{
Loss D.,  DiVincenzo D. P. and  Grinstein G. {\it Phys.  Rev.  
Lett.}  {\bf 69}, 3232, (1992).}

\def\refII{  von Delft J.
and  Henley C. L. {\it Phys. Rev. Lett.} {\bf 69}, 3236 (1992).}

\def\refIII{
 Garg A. {\it Europhys. Lett.} {\bf 22}, 205 (1993).}

The author has shown \cite{refIV} that mesoscopic magnets, in a 
suitably directed magnetic field, can exhibit {\it intermediate spin}.  
Intermediate spin values occur naturally \cite{refV} in the theory of 
anyons and are usually associated with the SO(2) group algebra which 
contains the single operator $\hat S_{z}$ with the spectrum $S_{z} = n 
- \alpha/2$, where $n$ an integer, and where $\alpha$ is the usual 
statistical parameter.  Evidently, anyons imply broken time reversal 
symmetry and, in the present context of a real physical magnet, the 
presence of a magnetic field.  The necessary reduction in symmetry 
from SO(3) (or SU(2)) to SO(2) (U(1)) will occur for large spin and an 
easy plane magnet.  For such a magnet, it has been shown \cite{refIV} 
the parity changes smoothly and periodically as a function of the 
magnetic field.  The absence of a tunnel splitting for half-integer 
points directly implies oscillations in the tunnel splitting, i.e., 
the oscillations of Garg are a trivial consequence of intermediate 
spin.

Very recently Wernsdorfer and Sessoli \cite{refVI} have experimentally 
observed double oscillations in the tunnel splitting for Fe$_{8}$ 
which is a molecular cluster with a net spin $S=10$.  Oscillations 
which occur with a field along the hard axis are interpreted in terms 
of intermediate {\it parity\/} (i.e., spin) while those which are 
induced by a field along the hard axis are discussed in terms of the 
Garg \cite{refIII} interference effect.  The principle purpose of this 
contribution is to show that both effects find a natural 
interpretation and a relatively simple quantitative description in 
terms of the author's \cite{refIV} approach to intermediate spin in 
magnets.  It will be shown, in the context of Fe$_{8}$, that periodic 
changes in parity can occur even for an easy ($y$-) axis magnet when 
the full SO(3) or SU(2) algebra is important.  However, it must be 
that the symmetry in the plane perpendicular ($x-y$-) plane is broken, 
so that it remains the case that there is an {\it easy\/} plane for 
tunneling between the classically defined minima.

\def\refIV{S. E. Barnes, J. Phys.  Cond.  Matter {\bf 10}, L665-L670 
(1998); cond-mat/ 9710302, (1997);  cond-mat/9901219 (1999).}

\def\refV{ See e.g., A. Khare {\it Fractional statistics and 
quantum field theory}, World Scientific, Singapore (1997).}

\def\refVI{W. Wernsdorfer and R. Sessoli, Science, {\bf 284}, 133 (1999).}

It has been shown by numerical simulation \cite{refVI} that Fe$_{8}$ 
is adequately described by the Hamiltonian:
\begin{equation}
{\cal H} = 
- D {S_{z}}^{2} + E \left( {S_{x}}^{2}- {S_{y}}^{2}\right)
+ g \mu_{B}\vec S \cdot \vec H
\label{un}
\end{equation}
where $D > E$ so that the $z$-axis is easy while the $x$-axis is hard.  
The term involving ${S_{y}}^{2}$ might be eliminated using 
${S_{y}}^{2}= S(S+1) - ({S_{x}}^{2}+ {S_{z}}^{2})$ so that, to within a 
constant, ${\cal H} = - (D - E) {S_{z}}^{2} +2 E {S_{x}}^{2}+ \vec S 
\cdot \vec h$ where $\vec h = g \mu_{B} \vec H$ is the field in energy 
units.

Formally this Hamiltonian has been considered in some detail in 
\cite{refIV} and by Preda and Barnes \cite{refVII} using the auxiliary 
particle method \cite{refVIII}.  This approach will be used for the 
detailed calculations.

\def\refVII{M. Preda and S. E. Barnes, J. Mag. Mag. Mat. in press (1999).}

\def\refVIII{S.~E. Barnes, J. Phys.  F{\bf 6}, 115; J. Phys.  F{\bf 
6} 1376 (1976); pg.  301, Proc.  XIX Congress Ampere, North Holland 
(1977); J. Phys.  F{\bf 7} 2637 (1977); J. de Physique, {\bf 39} 
C5-828 (1978); Advances in Physics {\bf 30} 801-938 (1981); J. Phys.  
C.M., {\bf 6}, 719-728 1994; S.~E. Barnes, B. Barbara, R. Ballou and J. Strelen, 
Phys.  Rev.  Lett.  {\bf 79}, 289 (1997); S. E. Barnes cond-mat/9710302 
(1979) and references therein.}

First consider a {\it longitudinal\/} field $h_{\ell}$, i.e., one 
along the $z$-axis, and to be specific {\it physical\/} whole integer 
spin value $S$.  The Hamiltonian only connects states $|S_{z} = S-n>$; 
$n$ an integer, for which $\Delta n $ is a multiple of two.  It 
follows that when the levels $S_{z} = -S$ and $S_{z} =S-n$ cross 
there can only be a tunnel splitting if $n$ is even.  The crossings at 
fields $h= s D$, with $s$ even correspond to whole integer physics 
while those with $s$ odd behave as if the spin was half-integer, i.e., 
a longitudinal applied field causes a periodic change of parity.

At a more formal level, following ref.  \cite{refIV}, it is observed 
that the longitudinal field $h_{\ell}$ can be eliminated by redefining the 
spin operator to be
\begin{equation}
\hat S_{z} \to \hat S_{z} - {\alpha \over 2}  ; \ \ \ \ 
\alpha = {h_{\ell} \over  D}.
\label{deux}
\end{equation}
which clearly has the intermediate spin spectrum $S_{z} = m - {\alpha 
\over 2}$; $m$ an integer.  This transformation does not change $[\hat 
S_{z},S^{\pm}] = \pm S^{\pm}$ {\it but\/} is not consistent with 
$[S^{+},S^{-}] = 2S_{z}$.  The argument that the problem maps to 
itself is more subtle than in ref.  \cite{refIV}.

The periodic behavior as a function of a {\it transverse\/} field can 
be understood in terms of the fixed point $E=D$.  Here, to within a 
constant, $\cal H$ reduces to $+2E{S_{x}}^{2} - S_{x}h_{t}$ where 
$h_{t}$ is the transverse field in energy units.  This defines another 
intermediate spin, or level crossing problem.  For an integer $S$, the 
$h_{t}=0$ ground state is non-degenerate, while it is doubly 
degenerate corresponding to a half-integer point when $h_{t}=2E$.  The 
``splitting'' is periodic with period $\Delta h_{t}=4E$.  This fixed 
point behavior explains qualitatively the tunnel splitting 
oscillations as a function of $h_{t}$, however a calculation of the 
general result for the period, the tunnel splitting, and the combined 
effects of longitudinal and transverse fields requires a detailed 
formalism and some tedious calculations.

In outline the calculation goes as follows.  The axis of quantization 
is put, $x \leftrightarrow z$, along the direction of the transverse 
field direction.  A basis $| S_z> \equiv |n>$ is chosen, and an 
auxiliary particle \cite{refVII,refVIII}, a fermion $f^{}_n$, is 
associated with each state via the mapping $|n> \to f^\dagger_n |>$ 
where $|>$ is a non-physical vacuum without any auxiliary particles.  
Defined is a bi-quadratic version of an operator $\hat O$ via: $ \hat 
O \to \sum_{n,n^\prime} f^\dagger_n <n|\hat O|n^\prime> 
f^{}_{n^\prime}$.  The constraint $ Q = \sum_n \hat n_n = \sum_n 
f^\dagger_n f^{}_n =1 $.  This procedure gives
\begin{eqnarray}
&{\cal H} =
 \sum_n \Big( (2E {n}^2 - h_{t}n) \nonumber\\
&-
{1\over 4} 
D[ \left( M_{n}^{n+1}\right)^{2} + 
\left(M_{n}^{n-1}\right)^{2}] \Big) f^\dagger_n f^{}_n\nonumber\\
&- 
{1\over 4} D \sum_n M_{n}^{n+1} M_{n+1}^{n+2} ( 
f^\dagger_{n+2} f^{}_n + f^\dagger_{n} f^{}_{n+2}  )\nonumber\\
&+
{1 \over 2} h_{\ell} \sum_n M_{n}^{n+1} 
( f^\dagger_{n+1} f^{}_n +  f^\dagger_{n} f^{}_{n+1} ),
\label{trois}
\end{eqnarray}
where the $M_{n}^{n+1} =[S(S+1) - n(n+1)]^{1/2}$ are the matrix 
elements of $S^{\pm}$.  Without a longitudinal magnetic field 
$h_{\ell}$, this is {\it two\/} isolated tight binding chains of 
spinless fermions $f^\dagger_n$.  A longitudinal field $h_{\ell}$ 
provides a coupling between these two chains.  The constraint $Q=1$ 
implies this is a single particle problem. Notice that the ``site'' 
indices are either whole of half-integer following the parity of the 
spin.

Schr\"odinger's equation 
\begin{eqnarray}
 &(\epsilon -  (2E {n}^2 - h_{t}n)) a_{n} = \nonumber\\
&  + {1\over4} 
D \Big[
M_{n}^{n+1} M_{n+1}^{n+2} a_{n+2}
+
M_{n}^{n+1} M_{n+1}^{n-2} a_{n-2}\nonumber\\
&+
[\left( M_{n}^{n+1}\right)^{2} + 
\left(M_{n}^{n-1}\right)^{2}]a_{n}
\Big] \nonumber\\
& -  {1 \over 2} h_{\ell}  ( M_{n}^{n+1} a_{n+1}+ M_{n}^{n-1} 
a_{n-1}){,}
\label{quatre} 
\end{eqnarray}
involves finite differences, where the wave-function $\Psi = \sum_{n} 
a_{n}f^{\dagger}_{n}|>$.  The difference in the site energies is 
$\sim 2E$ near the center $n \sim 0$ of the chains while the ``hopping 
term'' is $\sim S^{2} D$.  For the problem of interest $D$ is positive 
and the ground state wave function does not change sign. 

The ground state is located near the center of the chains and for most 
purposes it will suffice to replace the $M_{n}^{n+1}$ by a constant 
matrix element $S$ whence the problem reduced to that of a (discrete) 
harmonic oscillator.  It might be expected that {\it continuum 
approximation\/} is valid since $S^{2} D \gg 2E$.  If, $h_{\ell}=0$, and 
if the passage is made from the discrete to a continuum problem the two 
chains are are identical and there is no tunnel splitting.  In this 
rather unusual representation of a tunneling problem it is the 
exponentially small difference in energy between the ground state on 
the even site chain (the absolute ground state) and that on the odd 
site chain which leads to the ``tunnel'' splitting.

However, in the absence of fields and if $E=0$ this equation must 
reproduce the eigenenergies $- D m^{2}$ where $m= -S -(S-1) \ldots 
S-1, S$ is an integer quantum number.  In particular the ground state 
is {\it exactly\/} doubly degenerate with $m=\pm S$ and again with one 
such state on each chain.  In order to obtain correctly this limit use 
is made of the exact result for $E=H=0$, namely
\begin{equation}
a^{\pm S}_{n}=
{1 \over 2^{S}} 
e^{\mp i\pi}
\left[
{(2S)!
\over 
(S- n)! (S+ n)! }
\right]^{1/2}
\label{cinq}
\end{equation}
where the two solutions correspond to $S_{z}=\pm S$.  For large spin 
$S$ this approximates well to
\begin{equation}
a^{\pm S}_{n}=
{1 \over 2^{S}} 
e^{\mp i\pi} e^{-n^{2}/2S}, 
\label{six}
\end{equation}
i.e., ignoring the factor $e^{\mp i\pi}$ the wave-function is that of 
a simple harmonic oscillator fairly well localized relative to the 
limits at $n = \pm S$.  It can be made to lie on one or the other 
chains by taking the sum or difference of these solutions.  The 
system does not ``see'' that $n$ is limited.  Adding the term $2 
E{n}^{2}$ causes the wave function to become slightly more localized.  
Since this additional potential is harmonic the wave function remains 
harmonic.

Consider the problem with $h_{\ell} = 0$ but with both $E$ and 
$h_{t}$ finite. The wave function is taken to have the form
\begin{equation}
a_{n}  
=
f(n) a^{\pm S}_{n}.
\label{sept}
\end{equation}
Substituting this into Schr\"odinger's equation gives at a sufficient 
level of approximation:
\begin{eqnarray}
&\epsilon f(n)
=
(2E n^{2} - h_{t} n) f(n)\nonumber\\
&+ {1\over 4} S^{2} D [f(n+2)+ f(n-2) - 2f(n)].
\label{huit} 
\end{eqnarray}
At this point the {\it intermediate spin\/} transformation is made, 
i.e., the field $h_{t}$ is absorbed into the term $2E n^{2}$ by the 
shift $n \to n - (h_{t}/4 E)$.  There is an apparent displacement of 
$\delta =(h_{t}/4 E)$.  The small change in energy is unimportant and 
will be dropped.  However the {\it real\/} displacement $d$ is 
different.  With $h_{t}=0$ to a good approximation, $f(n) \sim 
e^{-(2E/D)^{1/2} n^{2}/2S}$ while for finite $h_{t}$, $f(n) \sim 
e^{-(2E/D)^{1/2} (n-(h_{t}/4E))^{2}/2S}$ so that $a_{n} \sim e^{- 
n^{2}/2S} e^{-(2E/D)^{1/2} (n-(h_{t}/4E))^{2}/2S}$ which is 
approximately $ e^{ (n-(h_{t}/2(2ED)^{1/2}))^{2}/2S}$, i.e., the {\it 
real\/} displacement of the net solution is
\begin{equation}
d = h_{t}/2(2ED)^{1/2},
\label{neuf}
\end{equation}
which corresponds to a statistical parameter $\alpha = 
h_{t}/(2ED)^{1/2}$.  The point to be made is that to a good 
approximation the solution for finite $h_{t}$ is the same as for 
$h_{t}=0$ {\it except\/} for a displacement of the site location from 
integer (or half-integer values for half-integer $S$) by $d$.

A Fourier transform, is made in order to examine the approach to the 
continuum limit and to generate a tunnel-barrier problem: $f(y) = 
{1\over \sqrt{2\pi}}\int dn\, e^{iny} a(n)$, whence Schr\"odinger's 
equation ($h_{\ell}=h_{t}=0$) is:
\begin{equation}
(2E {d^{2} \over d y^{2}}) f(y) = - 
{S^{2}D\over 2} [\cos 2 y - 1] f(y),
\label{dix}
\end{equation}
which is Mathieu's equation\cite{refIX}. The potential is periodic with 
period $\Delta y = \pi$ and the solutions form bands.  For a 
given energy, a solution might be characterized by a wave-vector $k$ 
and Floquet's (Bloch's) theorem\cite{refIX} implies that solutions are of 
the form $f_{k}(y) = e^{iky}u_{k}(y)$ where $u_{k}(y) = u_{k}(p+ 
\pi)$. 

\def\refIX{G. Blanch in {\it Handbook of Mathematical 
Functions}, eds.  M. Abramowitz and I. A. Stegun, U.S. Printing 
Office, Washington, D.C.}

Since the $k$-space is reciprocal to the $y$-lattice it is roughly 
equivalent to the real $x$-space and since this latter is discrete, 
making an inverse Fourier transform selects a particular wave vector 
$k$.  Also a shift in $k$-space is fully equivalent to a shift in real 
space.  Because of the intermediate spin shift $d$, the sites are 
precisely shifted from the whole or half integers.  Specifically for a 
whole integer spin, the solution which is finite on the {\it even\/} 
site chain implies $k=d$.  Similarly for the {\it odd\/} site chain 
$k=d-1$.  The {\it only\/} difference for a half-integer spin is an 
addition shift of $1/2$.  Without a intermediate spin (i.e., 
statistical shift) this makes the two $k$-values equivalent at $k = \pm 
1/2$ and directly implies a zero splitting in the absence of fields.

In the $y$-space, around $y=0$, $ f(y) = (1/ \sqrt{\beta \sqrt{\pi}}) 
e^{-y^{2}/2\beta^{2}}$ with $\beta^{2} = S \sqrt{2E/ D}$, and the 
nominal ground state energy is $(\omega_{0}/2)$; $\omega_{0} = 2S 
\sqrt{2ED}$.  The band energies are: $\epsilon_{k} = (\omega_{0}/2) + 
(w/2)\cos \pi k$ where the width \cite{refIX} $w = 8 \sqrt{2/\pi} 
\omega_{0}S_{f}^{1/2}e^{-S_{f}}$ and where the action $S_{f} = 2 
S\sqrt{(D/2E)}$.  The result for the ``tunnel splitting'', i.e., the 
difference in energy between the ground and first excited states is
\begin{eqnarray}
\delta E &=& 4 \sqrt{2 \over \pi} \omega_{0}S_{f}^{1/2}e^{-S_{f}} 
\left| \cos ( \pi d)\right| ;\nonumber\\
 S_{f} &=&  
2 S\sqrt{(D/ 2E)}{.}
\label{onze}
\end{eqnarray}
This result for the tunnel splitting is very similar to that for an 
easy plane magnet \cite{refI,refVII}.  The most significant difference 
is that the period for the oscillation, with a field along the hard 
axis is $2(2ED)^{1/2}$ rather than, in the present notation, $4E$.  
For $E$ small this result for the period agrees with the result of Garg 
$2(2E(D+E))^{1/2}$ quoted by Wernsdorfer and Sessoli \cite{refVI}.

The last step is to account for the longitudinal field $h_{\ell}$.  
There is a level crossing for the values $h_{\ell} = s D$; $s$ an 
integer.  Since they reflect the essential physics only results for 
these crossing points will be presented.  Consider a whole integer 
physical spin $S$, following the earlier discussion there are two 
types of crossing corresponding to the value of $s$.  For $s$ even, 
including $s=0$, the system behaves as if it is whole integer while 
for $s$ odd it has the characteristics associated with a half-integer 
total spin.  The proof goes as follows: The zero order wave functions 
for the states which cross are, e.g., $a^{- S}_{n}$ and $a^{S-s}_{n}$ 
which are both exact solutions.  From Eqn.  (\ref{cinq}): $a^{- S}_{n} 
= (-1)^{n}a^{S}_{n}$ where $a^{S}_{n}$ is even in $n$, while 
$a^{S-s}_{n}$ is even or odd as $s$ is even or odd.  Given there is 
tunneling, the appropriate wave functions are of the form, for large 
enough $S$:
\begin{equation}
a_{\pm , n} 
=
f(n) \left[
(-1)^{n} a^{ S}_{n} \pm a^{S-s}_{n}
\right].
\label{douze}
\end{equation}
If this function is substituted in Eqn.  (\ref{quatre}) the terms 
involving $h_{\ell}$ cancel to O($S$), so that the resulting 
difference equation and solution for $f(n)$ are unchanged.  However, 
the function $a_{\pm , n}$ is quite different depending on the even or 
odd character of $s$.  With $f(n)=1$, because of the term $(-1)^{n}$, 
$a_{\pm , n}$ jumps wildly between sites {\it but\/} defines a smooth 
function on {\it either\/} the even, $a^{e}_{\pm ,n}$, or odd , 
$a^{o}_{\pm ,n}$, sites.  When $s$ is odd it is {\it almost\/} the 
case that $a^{e}_{\pm ,n}= - a^{o}_{\pm ,-n}= -a^{e}_{\mp ,n} $.  {\it 
In fact\/}, if the smooth function which passes through the lattice 
points is displaced by $d = \delta n = 1/2$, then these relationships 
are exact.  As discussed above, it is just such a displacement with 
converts the whole into the half integer problem and demonstrates 
that the $s$ odd crossings are equivalent to a physical half-integer 
spin $S-s/2$.  When $s$ is even the functions $a^{e}_{\pm ,n}$ and 
$a^{o}_{\pm ,n}$ are very different, however (to within a sign) 
$a^{e}_{\pm ,n} = a^{o}_{\mp ,n}$ so that difference between the two 
solutions $a_{\pm , n}$ is a displacement by one lattice site or $d=1$ 
which is just the generalization of the same result for $s=0$ and 
which results in a the maximal tunnel splitting with $h_{t} = 0$. 
The tunnel splitting reflects the decrease in effective spin in the 
sense that it increases strongly with $s$.

% \vfill\eject

%-------------------References------------------------

\end{document}